# Direct observation of local antiferroelectricity induced phonon softening at a SrTiO$_3$ defect

Bo Han[1,2], Ruochen Shi[1,2], Huining Peng[3], Yingjie Lv[3], Ruishi Qi[2,4], Yuehui Li[1,2], Jingmin Zhang[2], Jinlong Du[2], Pu Yu[3], Peng Gao[1,2,5,6]*

[1] International Center for Quantum Materials, School of Physics, Peking University, Beijing 100871, China

[2] Electron Microscopy Laboratory, School of Physics, Peking University, Beijing 100871, China

[3] State Key Laboratory of Low Dimensional Quantum Physics and Department of Physics, Tsinghua University, Beijing 100084, China

[4] Physics Department, University of California at Berkeley, Berkeley, CA 94720, USA

[5] Collaborative Innovation Center of Quantum Matter, Beijing 100871, China

[6] Interdisciplinary Institute of Light-Element Quantum Materials and Research Center for Light-Element Advanced Materials, Peking University, Beijing 100871, China

* Corresponding author. E-mail: p-gao@pku.edu.cn

Defects in oxides usually exhibit exotic properties that may be associated with the local lattice dynamics. Here, at atomic spatial resolution, we directly measure phonon modes of an antiphase boundary (APB) in $SrTiO_3$ freestanding membrane and correlate them with the picometer-level structural distortion. We find that the $SrTiO_3$ APB introduces new defect phonon modes that are absent in bulk $SrTiO_3$. These modes are highly sensitive to the subtle structure distortion, i.e., the $SrTiO_3$ APB generates the local electric dipoles forming an antiferroelectric configuration, which significantly softens the transverse optical (TO) and longitudinal optical (LO) modes at $\Gamma$ point. Correlating the local phonons with the subtle structural distortion, our findings provide valuable insights into understanding the defect properties in complex oxides and essential information for their applications such as thermoelectric devices.

Interfaces and defects in the complex oxides such as SrTiO$_3$ usually exhibit exotic properties absent in the bulk matrix, including interfacial two-dimensional electron gas [1], superconductivity [2], conducting channel [3, 4], local flexoelectricity [5]. These properties originate from the changes of element compositions, strain and/or strain gradient [5], or oxygen octahedron distortion [6] in the vicinity of interfaces and defects due to the discontinued lattice. The deeper reason is the generation of new electronic structures and/or phonon modes that are localized at the interfaces and defects, and strongly associated with their specific atomic arrangements. Although the former (new electronic states of interfaces and defects) has been well recognized, the later (new phonon modes) has been relatively rarely discussed. Taking the thermal properties as an example, since phonons are primary heat carriers in oxide semiconductors and insulators, the phonon-scattering at a boundary in SrTiO$_3$ could be modulated and further induce unique thermal properties like sizeable thermal Hall conductivity [7]. For electrical properties, the electron-phonon interaction is the pivotal origin of the exotic properties in semiconductors and insulators. First principles calculations [8] and Raman experiments [9, 10] indicate that the oxygen vacancy defect in SrTiO$_3$ can induce extra Ti-O vibrational modes, and reduce the dielectric constant of SrTiO$_3$ thin films [11, 12]. At a SrTiO$_3$-LaAlO$_3$ hetero-interface, the electron-phonon coupling is enhanced and thus leading to its superconductivity [13]. Moreover, the soft phonon modes can give rise to ferroelectricity in epitaxy SrTiO$_3$ thin films due to the interfacial strain [14].

Although plenty of evidence has implied that the variations of atomic structure and corresponding vibrational modes at the interfaces and defects play crucial roles on their physical properties, their relations have been far from clear. It is challenging to directly reveal the phonon structure of a single interface or defect, because it requires atomic spatial resolutions for measuring both atomic arrangements and lattice vibrations, which is commonly not available for traditional optical methods. However, the recent advances in electron energy loss spectroscopy (EELS) equipped with monochromator in an aberration-corrected scanning transmission electron microscope (STEM) have made it possible to probe the vibrational properties at atomic scale for a single interface and defect [15, 16, 17, 18, 19, 20, 21], allowing us to study the structure-dependent lattice dynamics for individual defects.

In this study, we combined such a STEM-EELS measurement with density functional theory (DFT) calculations to correlate the phonon structure of an antiphase boundary (APB) in the prototype perovskite $SrTiO_3$ with its picometer-level structure distortion. We found that the edge sharing oxygen octahedrons at the $SrTiO_3$ APB generates new phonon modes with the energy at ~48 meV and ~75 meV. We also observed that local electric dipoles (~10 pm in displacement) emerge at the APB, constituting an antiferroelectric (AFE) configuration. Moreover, we clarified that the energy values of defect phonon modes strongly rely on such local subtle structural distortions, i.e., the AFE distortions cause defect phonon softening compared to the distortion-free case. Our study provides unambiguous correlation of vibration properties with the subtle atomic distortion for a specific defect in $SrTiO_3$, which helps us to understand its unique properties such as reduced thermal conductivity in thermoelectric $SrTiO_3$.

We used high-angle annular dark-field (HAADF) imaging combined with energy dispersive X-ray spectroscopy (EDS) to reveal the atomic structure of $SrTiO_3$ APB. The sample was fabricated into a free-standing $SrTiO_3$ membrane to probe the intrinsic defect structure, independent of the epitaxial substrate (See Supplementary materials for experimental details). From the atomically resolved HAADF image in Fig. 1(a), the boundary contains two closely aligned Ti columns (viewing from [001]). The atomically resolved elemental maps in Fig. 1(b) also verify the connection of two $(TiO_2)^0$ planes at the SrTiO$_3$ boundary. Viewing from [001] direction, the measured distance between the two interfacial Ti columns (~1.96 Å, detailed in Fig. S1) is coincidently about the half of the lattice constant of bulk SrTiO$_3$ (PDF#35-0734, $a = b = c = 3.905$ Å), indicating an edge sharing configuration of TiO$_6$ octahedron along [001] direction [22]. As depicted in Fig. 1(c-d), edge sharing TiO$_6$ octahedrons connect two adjacent $(TiO_2)^0$ planes, forming the APB with a half-unit-cell offset. The energy-loss near-edge structures of Ti-*L* edge in Fig. 1(e-f) exhibit a less peak separation, indicating the change of crystal field and an increase in Ti$^{3+}$ component at the APB. This is originated from edge sharing oxygen octahedrons, i.e., a lower O/Ti ratio compared to the bulk matrix. Such a Ti-rich APB is distinct from well-known Ruddlesden-Popper planar faults with $(SrO)^0$ planes connected in perovskite structure [23]. Its formation may stem from the surface steps of the substrate [24], and/or the merging of different domains during nucleation and growth [25].

For APBs, the unique atomic configurations change the bonding environment and thus the phonon structure. In order to study the localized defect phonons, the vibrational EELS across the SrTiO$_3$ APB are displayed in Fig. 2(a), which were performed using a dark-field EELS geometry (portrayed in Fig. S2) to minimize the delocalization effect [17, 26]. The transverse optical (TO) modes and the longitudinal optical (LO) modes in bulk SrTiO$_3$ region contribute three dominating peaks on its vibrational EELS, containing LO3 (~99 meV), LO2/TO3 (~65 meV) and TO1/TO2 (~20 meV). Intriguingly, at the APB, the vibrational EELS show extra enhanced signals in the range of 40–53 meV and 65–85 meV, located in the gaps between LO3, LO2/TO3 and TO1/TO2, which are also shown by the spectral difference in Fig. 2(b). These in-gaps signals imply the existence of defect modes that are absent in the bulk matrix. They can be verified by the theoretical calculations using density functional perturbation theory (DFPT). From the calculated density of state (DOS) spectra in Fig. 2(c-d), the additional defect phonon DOS emerges at the boundary, resulting in the EELS intensity increase in the gaps between LO3, LO2/TO3 and TO1/TO2. From the phonon intensity maps in Fig. 2(e-f), these in-gaps phonon modes (at ~48 meV and ~75 meV) correspond to the so called "interfacial phonon modes" in previous study, i.e., they are localized at APBs and are absent in bulk materials [27]. The simulated phonon intensity maps based on DFPT [Fig. 2(g-h)] also show intensity increase at the APB, which are in good agreement with the experimental observations.

In order to explore the origin of the defect phonons, the subtle structure distortion at the $SrTiO_3$ APB is considered. At APBs, the repulsive $TiO_6$ octahedrons account for the increased Ti-Ti displacement and create antiparallel dipoles [Fig. 3(a)], forming an AFE configuration at unit-cell scale. As illustrated in Fig. 3(b), the antiferroelectricity at APB can be described by the displacements between $d_1$ and $d_2$, where $d_1$ is defined by the experimental Ti-Ti distance and $d_2$ is the "ideal" Ti-Ti distance. At the APB, the "ideal" $d_2$ is the half of the $SrTiO_3$ lattice constant. The mapped $d_1$-$d_2$ displacement in Fig. 3(c) shows an increased Ti-Ti displacement with the maximum up to ~17 pm, and an average of ~10 pm in Fig. 3(d). Such a displacement separates the center of positive and negative charges, leading to polarization at the APB. From Fig. 3(e), the inverse directions of polarizations at the defect coincide with the defect antiferroelectricity of $SrTiO_3$ APB.

In fact, such a tiny structural distortion of $TiO_6$ octahedrons at $SrTiO_3$ APB, give rise to significant change in its phonon structure. Fig. 4(a) present the calculated phonon dispersion curves of an "ideal" APB (without Ti-Ti AFE distortions), of which the momentum path in Brillouin Zone is illustrated in Fig. S3. The distortion-free $TiO_6$ octahedrons predominantly contribute defect modes at 83 meV and 53 meV. From the phonon dispersion curves of the APB with AFE distortion, Ti-Ti displacements can soften the defect modes, as the blue arrow highlighted in Fig. 4(b). The representative defect Mode I in Fig. 4(c) is softened from 83 meV ("ideal" APB) to 75 meV ("AFE" APB). Mode II in Fig. 4(d) is softened from 53 meV ("ideal" APB) to 43 meV ("AFE" APB). From the eigenvectors of Mode I and Mode II, the vibration enhancements are locally concentrated at the APB and they are absent in the bulk region, showing evident defect-confined feature in real space. For comparison, the phonon vibration of bulk $SrTiO_3$ mainly occurs in the bulk region, as shown in Fig. 4(e). In momentum space, the defect phonons mainly come from $\Gamma$ point, which is also verified by the phonon dispersion and the momentum resolved scattering cross section in Fig. S4.

Defect phonons localized at the defect are expected to have a profound influence on the physical properties of $SrTiO_3$. $SrTiO_3$ possess excellent thermal stability for thermoelectric performances [28], for which large Seebeck coefficient, high electrical conductivity and low thermal conductivity are highly desirable. Previous study showed that APBs in $SrTiO_3$ cause negative thermal expansion [29], and construct barriers for ion migrations [30], which are beneficial for the thermoelectric applications [31]. In our study, the phonon scattering is enhanced at APB due to the remarkable energy mismatch between the defect phonons and the bulk phonons. As a result, the $SrTiO_3$ APB act as barrier for phonon transmission through defect phonon modes, thus may reduce the thermal conductivity of $SrTiO_3$.

In summary, we used vibrational STEM-EELS to probe the defect phonon of a $SrTiO_3$ APB. We found that the APB possesses localized defect phonon modes which are absent in the bulk. The defect phonon modes are most prominent near the $\Gamma$ point and softened due to defect antiferroelectric distortions. Such defect modes may enhance the phonon scattering and further yields good thermoelectric performance. Our study correlates the phonon structure with the subtle structure distortion of a single defect in a complex oxide, shedding light on the defect engineering for thermoelectric devices.

**Author Contribution**

P.G. conceived and supervised the project; B.H. performed the STEM-EELS experiment and data analysis assisted by R.C.S., R.S.Q., and Y.H.L. with the guidance of P.G.; R.S.Q., R.C.S., and Y.H.L. developed ab initio vibrational EELS simulation and assisted B.H. with the DFT calculation; R.S.Q. and R.C.S. designed the toolbox for data processing; H.P. fabricated $SrTiO_3$ freestanding membranes. Y.L. grew the samples with the guidance of P.Y.; J.Z. and J.D. helped the data interpretation. B.H. wrote the manuscript under the direction of P.G; All the authors contributed to this work through useful discussion and/or comments to the manuscript.

**Acknowledgement**

The work was supported by the National Key R&D Program of China (2021YFA1400503), the National Natural Science Foundation of China (52125307, 11974023, 52021006, T2188101), the "2011 Program" from the Peking-Tsinghua-IOP Collaborative Innovation Center of Quantum Matter, Youth Innovation Promotion Association, CAS. The sample preparation at Tsinghua was supported by the National Basic Research Program of China (Grant No. 2021YFE0107900); the NSFC (grants 52025024 and 51872155) and the Beijing Nature Science Foundation (grant No. Z200007). We acknowledge Electron Microscopy Laboratory of Peking University for the use of electron microscopes. We acknowledge High-performance Computing Platform of Peking University for providing computational resources for the DFPT calculation.

# Figures and captions

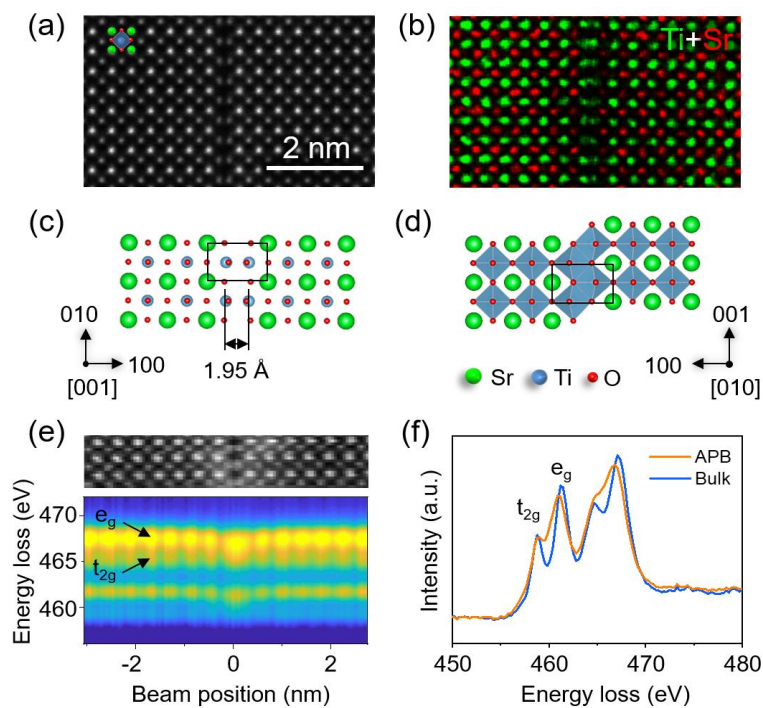

Figure 1. The atomic structure of a SrTiO$_3$ anti-phase boundary (APB). (a) A HAADF image of an APB in SrTiO$_3$, overlapped with an atomistic model. Green (Sr), blue (Ti) and red (O) spheres. (b) The EDS maps of Ti (green) and Sr (red) showing cationic distribution in the vicinity of the boundary. (c-d) The atomistic schematics of the APB, viewing from (c) [001] and (d) [010] direction, respectively. At the APB, TiO$_6$ octahedrons are edge sharing along [001] direction. (e) The Ti-*L* edge core loss EELS across the APB. (f) The Ti-*L* edge spectra extracted from APB (orange) and bulk (blue) showing the changed crystal-field at the APB.

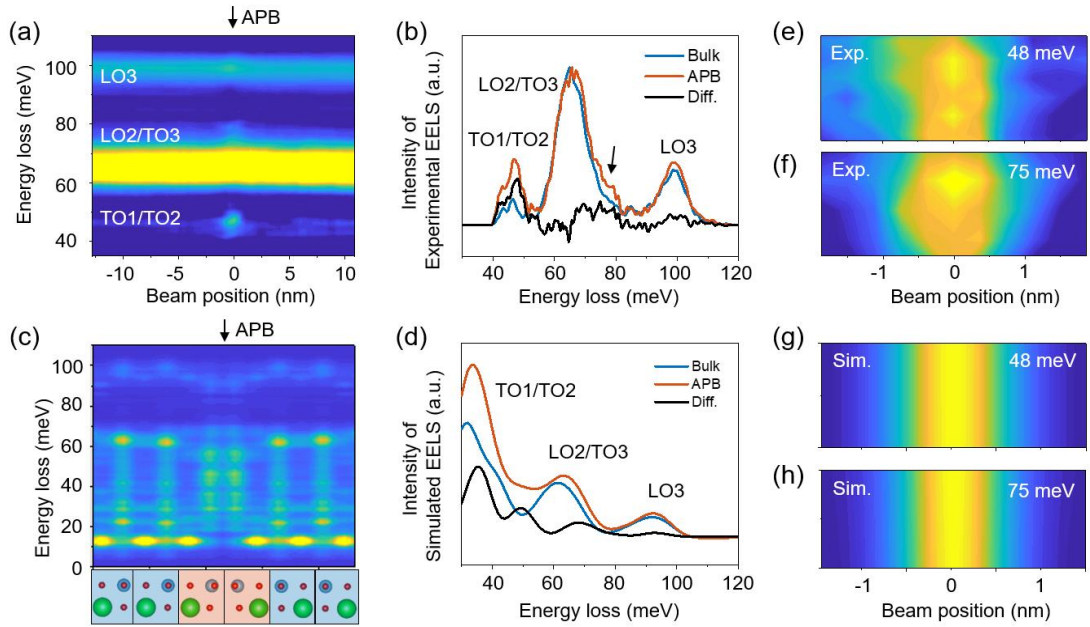

Figure 2. The localized defect phonon imaging at the SrTiO$_3$ APB. (a) The experimental phonon spectra across the SrTiO$_3$ APB. (b) Phonon spectra extracted from bulk (blue), interface (orange), and their difference (black). (c) The atomically resolved phonon density of state (DOS) across APB based on DFT calculations. The distorted TiO$_6$ octahedrons at APB contribute to defect DOS. (d) The simulated phonon spectra extracted from bulk (blue), interface (orange), and their difference (black). The intensity of the spectrum increases at the APB. (e-f) The experimental phonon intensity map at (e) 48 meV and (f) 75 meV. (g-h) The simulated phonon intensity map at (e) 48 meV and (f) 75 meV.

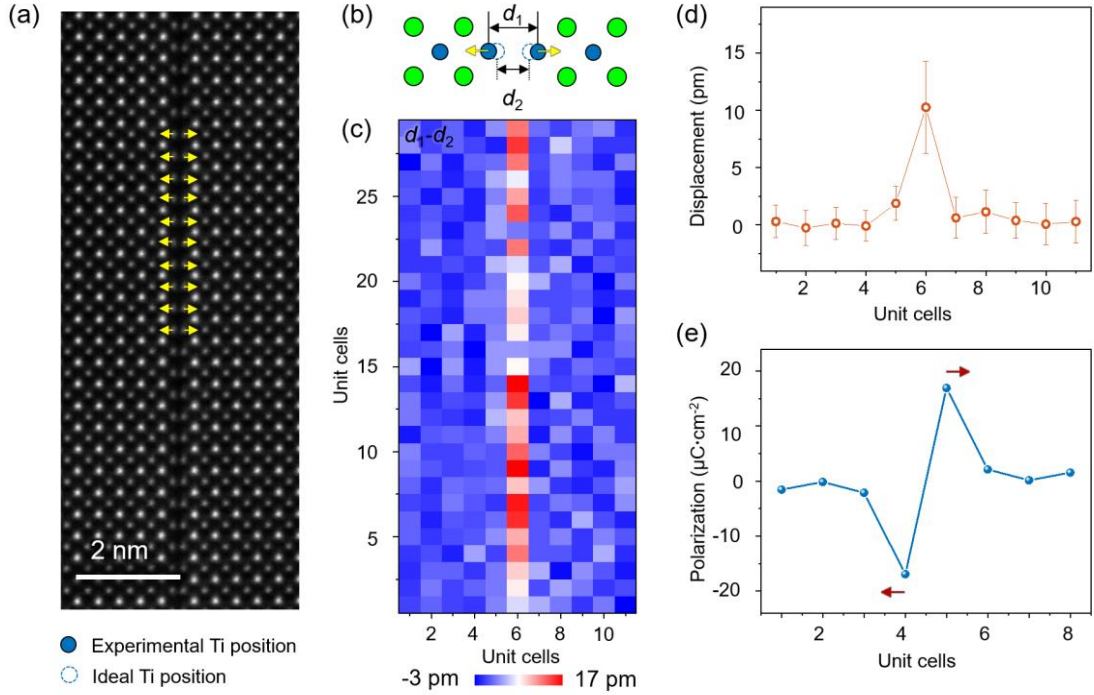

Figure 3. The antiferroelectricity at SrTiO$_3$ APB. (a) The experimental HAADF image. Yellow arrows: the displacement directions of Ti cations respect to their ideal positions. (b) The atomistic schematic of APB viewing from [001] showing the increased Ti-Ti distance in the projection. $d_1$: the experimental Ti-Ti distance; $d_2$: the ideal Ti-Ti distance (corresponding to perfect bulk SrTiO$_3$). (c) The mapping of $d_1 - d_2$ across the APB. (d) The plot of $d_1 - d_2$ based on HAADF images. The displacements ($d_1 - d_2$) increase at APBs. (e) The calculated polarization across the APB. Red arrows: the polarization directions at APB are opposite.

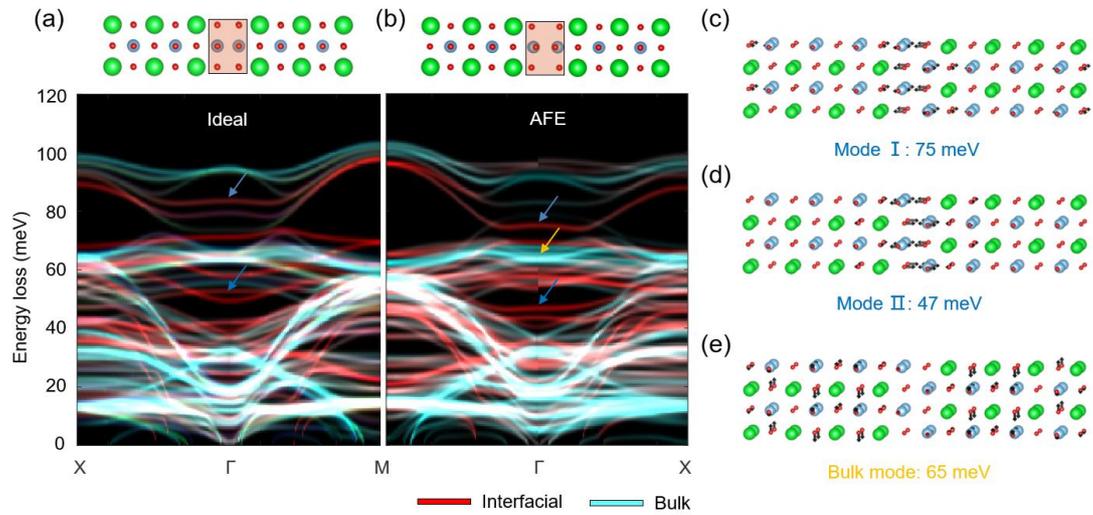

Figure 4. The origin of SrTiO$_3$ defect phonon softening at the APB. (a-b) The atomic structure of APB (a) without distortion, (b) with AFE distortion, and their corresponding phonon dispersion curves. Red curves: defect components, cyan curves: bulk components. (c-d) The eigenvectors showing the defect modes of the "AFE" APB at (c) 75 meV and (d) 47 meV. (e) The eigenvectors showing a bulk phonon mode of SrTiO$_3$ at 65 meV.

**Supplementary materials:**

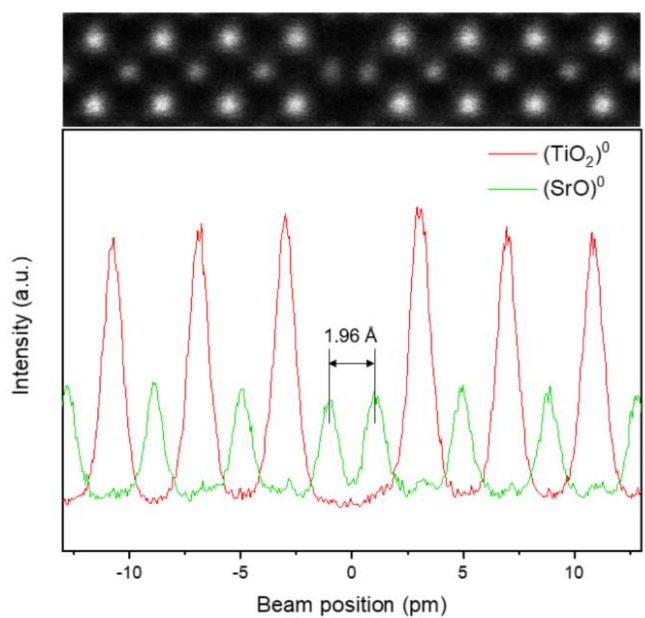

**FIG. S1.** The measured atomic distance at SrTiO$_3$ APB

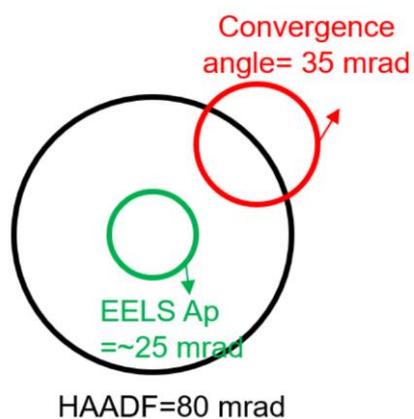

**Figure. S2.** Set-up of the dark-field EELS for interfacial acquiring. The 35 mrad probe is deliberately shifted about 80 mrad away from the EELS entrance aperture (25 mrad), thus selectively collecting electrons scattered to a high angle. Such a dark-field vibrational EELS set-up can suppress the delocalized dipole scattering and facilitate the detection of localized phonon even at atomic resolution.

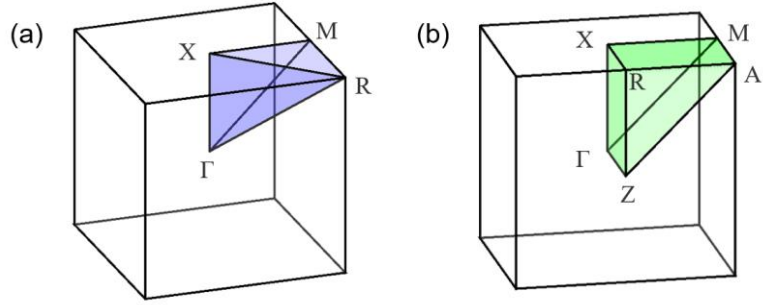

**Figure. S3.** The Brillouin Zone of (a) the cubic bulk SrTiO$_3$ and (b) the tetragonal interfacial model.

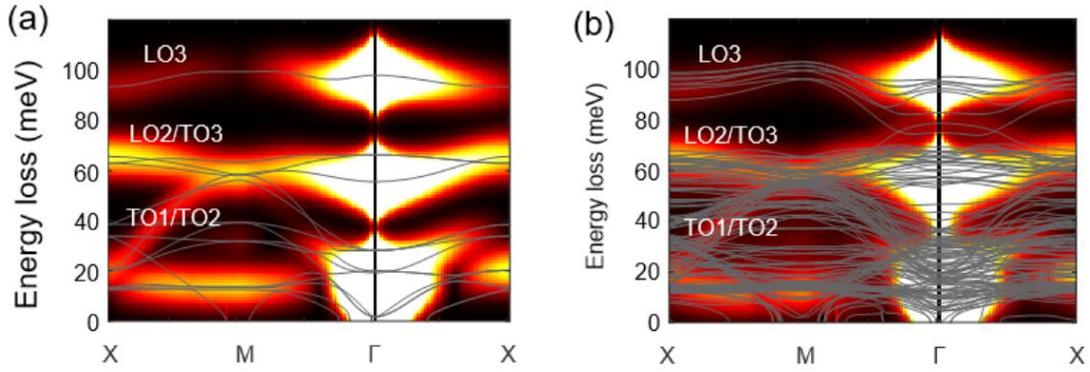

**Figure. S4.** The momentum resolved scattering cross section of (a) cubic bulk SrTiO$_3$ and (b) SrTiO$_3$ APB.

In bulk SrTiO$_3$ [Figure. S4(a)], LO3 and LO2/TO3 dispersion curves along Γ–X–M–Γ–R are flat, accounting for two vibrational EELS peaks at ~65 meV and ~99 meV. Notably, there is a gap between these two main branches, i.e., there is no phonon mode and EELS peaks between ~65 meV and ~99 meV. However, at SrTiO$_3$ APBs [Figure. S4(b)], a noticeable additional interfacial branch appears between LO3 and LO2/TO3. This indicates that the enhancement of EELS intensity between 65 meV and 99 meV at APB comes from the extra interfacial modes in the vicinity of the Γ point.

## Experimental Methods

**Thin film heterostructure preparation.** Heterostructures of $SrTiO_3/SrCoO_{2.5}$ thin films on $(LaAlO_3)_{0.3}$–$(SrAl_{0.5}Ta_{0.5}O_3)_{0.7}$ (LSAT) substrate were grown by pulsed laser deposition method. The $SrCoO_{2.5}$ layers were grown at a temperature of 750 °C under an oxygen partial pressure of 13 Pa, with laser (KrF, $\lambda = 248$ nm) energy density of 1.1 J/cm$^2$ and frequency of 2 Hz. The $SrTiO_3$ layers were gown under an oxygen partial pressure of 13 Pa, with a substrate temperature of 700 °C, and the laser energy density of 1.3 J/cm$^2$ and repetition rate of 2 Hz. After the deposition, samples were cooled down to room temperature at growth pressure with the cooling rate of 10 °C per minute. The sample thickness was controlled by the growth time at calibrated growth rate, and then further confirmed through the X-ray reflectometry measurements.

**Fabrication of $SrTiO_3$ freestanding membranes.** The heterostructure of $SrTiO_3$/$SrCoO_{2.5}$/LSAT was immersed into 36% acetic acid solution at room temperature to dissolve the $SrCoO_{2.5}$ sacrificed layer. After dissolving the $SrCoO_{2.5}$ layer within a few hours, the freestanding $SrTiO_3$ nanomembrane was exfoliated from the substrate and suspended in solution. Then, a TEM Cu grid was used to put $SrTiO_3$ nanomembrane out and the freestanding nanomembrane was adhered on the grid. The whole etching and exfoliation processes were performed in ambient conditions.

**HAADF EDS characterizations.** The atomically resolved STEM-HAADF and EDS images were acquired using an aberration-corrected Titan Themis G2 microscope at 300 kV with a beam current of ~80 pA, a convergence semi-angle of 30 mrad, and a collection semi-angle snap in the range of 39–200 mrad.

**EELS characterizations and processing.** The STEM-EELS were recorded using a Nion HERMES 200 microscope with both monochromator and the aberration corrector operating at 60 kV at room temperature. The convergence semi-angle was 35 mrad and the collection semi-angle was 24.9 mrad. The dwell time was 1600 ms/pixel, the dispersion was 0.5 meV/ch, and the EELS camera setting was 260×2048, i.e., the EEL spectra recorded for each pixel is summed by 260 channels on EELS camera. The EELS background was fitted by a Pearson function [16] and then subtracted. Lucy-Richardson deconvolution was then employed to ameliorate the broadening effect caused by finite energy resolution.

**Ab initio calculations.** We used ab initio DFPT to investigate the interfacial mode of $SrTiO_3$ boundaries. The DFPT calculation were performed within Quantum ESPRESSO [32, 33] using the projector augmented wave (PAW) pseudopotential [34] and the modified Perdew-Burke-Enzerhof (PBEsol) exchange-correlation functional, which gives reliable lattice parameters and phonon frequencies for bulk $SrTiO_3$ [35]. The supercell model of $SrTiO_3$ boundary was constructed using two slices of bulk $SrTiO_3$, each of which contains 4 layers (unit-cells) and connected by their $(TiO_2)^0$ termination planes. The atomic positions of the supercell were optimized under a tetragonal symmetry (with optimized cell parameters $a = b = 3.895$ Å, $c = 32.122$ Å) until the atomic force on each atom is within $1 \times 10^{-4}$ Ry/Bohr. In the geometric optimizations, the cutoff energy is 600 eV. For self-consistent field calculations, the charge cutoff energy is 600 eV and the wavefunction cutoff is 60 eV. The phonon dispersion and DOS were calculated by interpolating the dynamic matrix on a 6×6×1 momentum mesh.

For the vibrational EELS simulation, the electron beam was modeled as a Gaussian beam. The scattering cross section between electrons and phonons were calculated according our previous work [20]. The dark-field vibrational EELS were simulated by summing the scattering cross section in the momentum space which corresponds to a convolution of the experimental collection aperture and the angle distribution of the incident electron beam.